# The Impact Failure Detector

Anubis G. M. Rossetto, Cláudio F. R. Geyer
Institute of Informatics,
Federal University of Rio Grande do Sul
Porto Alegre, Brazil
{agmrossetto, geyer}@inf.ufrgs.br

Luciana Arantes, Pierre Sens
Laboratoire d'Informatique de Paris 6 (LIP6)
Université Pierre et Marie Curie, CNRS, INRIA
Paris, France
{luciana.arantes, pierre.sens}@lip6.fr

*Abstract*— **This work proposes a new and flexible unreliable failure detector whose output is related to the trust level of a set of processes. By expressing the relevance of each process of the set by an impact factor value, our approach allows the tuning of the detector output, making possible a softer or stricter monitoring. The idea behind our proposal is that, according to an acceptable margin of failures and the impact factor assigned to processes, in some scenarios, the failure of some low impact processes may not change the user confidence in the set of processes, while the crash of a high impact factor process may seriously affect it. We outline the application scenarios and the proposed unreliable failure detector, giving a detailed account of the concept on which it is based.**

## I. Introduction

An unreliable failure detector (FD) is a basic service that can enable the development of fault tolerant distributed systems [1]. A FD can be seen as an oracle that gives information, not always correct, about processes failures [2]. The majority of them are based on a binary model, where monitored processes are either "trusted" or "suspected". Therefore, many of the existing FD, such as those defined in [1] [3], output the set of processes that is currently suspected to have crashed. A non binary approach is present in [4], the accrual failure detector, which outputs a suspicion level on a continuous scale.

This paper presents a new unreliable failure detector, called **Impact**. The Impact FD outputs a trust level concerning a set of monitoring processes. The output can be considered as the degree of confidence in this set of processes. To this end, an *impact factor* is assigned to each process of the set. Furthermore, a *fault margin* parameter defines an impact factor limit value, below which the confidence degree on the set is not affected. Hence, depending on the trust level output, the users can decide whether it is reliably operating or if some measures may be taken (urgently or not, with regard to the trust level output).

**Motivation**: The Impact FD can be applied to different distributed scenarios and it is flexible enough to meet different needs. It is quite suitable for environments where there is node redundancy. The following two examples illustrate scenarios where the Impact detector can be used.

A system in the area of healthcare requires the use of several sensors to measure different kinds of information about the health status of a person, such as, vital signs, location, falls, gait patterns, and acceleration. From this perspective, this scenario is critical since faults in the components can put the patient at serious risk. For instance, we can consider a scenario with four sensors: $q_1$ - *body temperature;* $q_2$ - *pulse;* $q_3$ - *electrocardiogram(ECG); and* $q_4$ - *galvanic skin* as well as an intermediate node responsible for collecting information from these sensors and taking appropriate action based on the output of the Impact FD. In this example, some sensors are not considered to be critical, such as the sensor $q_1$ which measures the temperature; however, others are extremely critical such as $q_3$, the ECG sensor. Therefore, the impact factor assigned to $q_3$ is higher than $q_2$'s. Furthermore, $q_2$ collects data on the heartbeats and $q_3$ on the electrical activity of the heart. However, $q_3$ is a type of sensor that also collects data on the heartbeats. Hence, since there is redundancy of information, i.e., the failure of $q_2$ sensor is not critical enough to make the system vulnerable and endanger the life of the monitored person. We could, therefore, define a *fault margin* equals to $q_2$'s impact factor. On the other hand, if $q_3$ fails, the confidence of the system is compromised.

A second example concerns Connected Dominating Set (CDS), which has been proposed as a virtual backbone of wireless ad hoc networks [5]. A dominating set (DS) of a graph G = (V, E) is a subset V' $\subset$ V such that each node in V - V' is adjacent to some node in V', and a connected dominating set (CDS) is a dominating set which also induces a connected subgraph. Nodes in V' are denoted *dominator nodes* (DN) and are the only responsible for relaying messages over the network. In such a scenario, we could think that every DN monitors the liveness of its neighbor nodes (the monitored set). However, since the failure of a DN may seriously affect the connectivity of the CDS, DN neighbors should have a higher impact factor than the other neighbors. In its turn, the *fault margin* could be set with a value that expresses that the connectivity of the network should always be guaranteed.

## II. The Impact Unreliable Failure Detector

The Impact FD can be defined as an unreliable failure detector that provides an output related to the trust level with regard to a set of processes. If the FD output is under a given limit value, defined by the user, the confidence in the set of processes is ensured.

We consider a distributed system that consists of a finite set of processes $\Pi = \{p_1,...,p_n\}$. The failure model is based on [1].

Failures are only by crash. A process can be correct or faulty, i.e., a process is faulty if it has crashed, otherwise it is correct. We assume the existence of some global time denoted T. A failure pattern is a function F: T → $2^\Pi$, where F(t) is the set of processes that have failed before or at time t. The function correct(F) denotes the set of correct processes, i.e., those that never belong to failure pattern (F), while faulty(F) denotes the set of faulty processes, i.e., the complement of correct(F) with respect to $\Pi$.

A process $p \in \Pi$ monitors $m$ processes $q$ ($q \in S = \{q_1,...,q_m\}$: $S \subset \Pi$). Each process $q \in S$ has a fault *impact factor* ($I_q$). Process $p$ knows its own identity and those of processes in S, as well as the impact factor of the latter. It does not necessary know $\Pi$. Furthermore, an acceptable margin of failures, denoted $fault\_margin_p^S$, characterizes the degree of failure acceptable flexibility of $p$ in relation to $S$. The value of every impact factor ($I_q$) must be within $0 < I_q \leq fault\_margin_p^S$. We should point out that both the *impact factor* and the $fault\_margin_p^S$ parameters render the estimation of the confidence of $S$ by $p$ more flexible. For instance, it might happen that some processes in $S$ are faulty or suspected of being faulty and p always trusts S.

We define *sum(set)* as the function that returns the sum of the impact factors of all processes in *set*. Note that for the monitoring set $S$, $sum(S)$ must be at least equal to the $fault\_margin_p^S$. We also define a trust limit, such that $trust\_limit_p^S = sum(S) - fault\_margin_p^S$.

We denote $trusted_p^S(t)$ as the set of processes in $S$ that $p$ considers not to be faulty at time $t \in T$. The Impact FD ($I_p^S$) outputs the trust level of $S$, which expresses the confidence that $p$ has in the set S:

**Definition 1** (trust level): The trust level at $t \in T$ of processes $p \notin F(t)$ with respect to the set of processes $S$ is the function $trust\_level_p^S$: T → R, such that $trust\_level_p^S(t) = sum(trusted_p^S(t))$.

If $trust\_level_p^S(t) > trust\_limit_p^S$, $S$ is considered to be trusted at $t$ by $p$, i.e., the confidence of $p$ in S has not been affected; otherwise $S$ is considered not trusted by $p$ at $t$.

Fig. 1 shows an example with a set of four processes. The impact factor was defined according to the relevance of each process of set $S$. The $fault\_margin_p^S$ is equal to 1. Thus, $sum(S)$ and $trust\_limit_p^S$ are respectively equal to 2.6 and 1.6 ($sum(S) - fault\_margin_p^S$). In this set, we can observe that process $q_1$ has low relevance and $q_3$ is highly relevant. Several situations are shown and the set $S$ is considered not trusted when the $trust\_level_p^S(t)$ is smaller than or equal to $trust\_limit_p^S$.

The approach of the Impact FD makes possible to carry out the monitoring in both softer and stricter way. Furthermore, if some properties are defined, some classes of Impact FD can also be defined. For instance:

**Prop. 1** (*Impact strong completeness $_p^S$*): If $p$, which monitors the set $S$, is correct, there is a time after which $p$ does not trust any crashed process of $S$.

```
∃t ∈ T, p ∈ correct(F), ∀q ∈ (faulty(F) ∩ S),
        ∀t' ≥ t, q ∉ trusted_p^S(t')
```

| S={$q_1,q_2,q_3,q_4$} | | | $I_{q1}$=0.2; $I_{q2}$=0.8; $I_{q3}$=1; $I_{q4}$=0.6; | |
|---|---|---|---|---|
| t | F(t) | $trusted_p^S(t)$ | $trust\_level_p^S(t)$ | Status |
| 1 | {$q_1$} | {$q_2, q_3, q_4$} | 2.4 | TRUSTED |
| 2 | {$q_1, q_2$} | {$q_3, q_4$} | 1.6 | NOT TRUSTED |
| 3 | {$q_4$} | {$q_1, q_2, q_3$} | 2 | TRUSTED |
| 4 | {$q_1, q_3$} | {$q_2, q_4$} | 1.4 | NOT TRUSTED |
| 5 | {$q_3$} | {$q_1, q_2, q_4$} | 1.6 | NOT TRUSTED |

*$fault\_margin_p^S$ = 1; sum(S) = 2.6; $trust\_limit_p^S$ = 1.6*

Fig. 1. Example of *p's* FD output related to *S*

**Prop. 2** (*Eventual impact strong accuracy $_p^S$*): If $p$, which monitors the set $S$, is correct, there is a time after which correct processes of $S$ are never suspected by $p$.

```
∃t ∈ T, ∀t' ≥ t, p, ∀q ∈ (correct(F) ∩ S),
        q ∈ trusted_p^S(t')
```

**Prop. 3** (Eventual set sum accuracy):

```
∃t ∈ T, p ∈ correct(F),
∀t' ≥ t, trust_level_p^S(t') = sum(correct(F) ∩ S)
```

**Eventual Perfect Impact FD** ($\Diamond IP_p^S$) **class**: For process $p$ and set $S$, Prop. 1, Prop. 2, and Prop. 3 are satisfied.

### III. CONCLUSIONS AND FUTURE WORK

In this paper, we have presented and defined a new unreliable failure detector, called Impact, that provides an output related to a set of processes and not just to each one individually. We have also described some scenarios suitable for the Impact FD. Both the *impact factor* and the *fault margin* provide a degree of flexible applicability, since they enable the user to tune the FD output in accordance with the specific needs of the application. They also might weaken the rate of false responses when compared to traditional unreliable failure detectors.

In a further step of this research, we intend to define other properties of the Impact FD and address failure detector reductions from the latter to other well-known detectors.

ACKNOWLEDGMENT

This work was partially supported by grant 012909/2013-00 from the Brazilian Research Agency (CNPq).